\documentclass[12pt]{article}
\usepackage{amsmath}

\csname @addtoreset\endcsname{equation}{section}
\newcommand{\eq}{\begin{equation}}
\newcommand{\feq}{\end{equation}}
\newcommand{\eqn}{\begin{eqnarray}}
\newcommand{\feqn}{\end{eqnarray}}
\newcommand{\arr}{\begin{eqnarray*}}
\newcommand{\farr}{\end{eqnarray*}}

\begin{document}

\begin{titlepage}
\begin{flushright}
IFUM 666/FT\\
CAMS/00-09\\
hep-th/0010200
\end{flushright}
\vspace{.3cm}
\begin{center}
\renewcommand{\thefootnote}{\fnsymbol{footnote}}
{\Large \bf Charged Rotating Black Holes in 5d Einstein-Maxwell-(A)dS Gravity}
\vfill
{\large \bf {D.~Klemm$^1$\footnote{email: dietmar.klemm@mi.infn.it} and
W.~A.~Sabra$^2$\footnote{email: ws00@aub.edu.lb}}}\\
\renewcommand{\thefootnote}{\arabic{footnote}}
\setcounter{footnote}{0}
\vfill
{\small
$^1$ Dipartimento di Fisica dell'Universit\`a di Milano
and\\ INFN, Sezione di Milano,
Via Celoria 16,
20133 Milano, Italy.\\
\vspace*{0.4cm}
$^2$ Center for Advanced Mathematical Sciences (CAMS)
and\\
Physics Department, American University of Beirut, Lebanon.\\}
\end{center}
\vfill
\begin{center}
{\bf Abstract}
\end{center}
We obtain charged rotating black hole solutions to the theory of
Einstein-Maxwell gravity with cosmological constant in five dimensions.
Some of the physical properties of these black holes are discussed.

\end{titlepage}

\section{Introduction}

There has been some renewed interest in the study of black hole physics in
gravitational theories with cosmological constant. Theories with negative
cosmological constant can be embedded in a supersymmetric setting obtaining
gauged supergravity theories in various dimensions. Gauged supergravities
admit anti-de~Sitter space as a vacuum state, and thus black hole solutions
of these theories are of physical relevance to the proposed AdS/CFT
correspondence \cite{adscft}. In particular, the study of AdS black holes
can give new insights into the nonperturbative structure of some conformal
field theories.

On the other hand, black holes in backgrounds with positive cosmological
constant, i.~e.~in de~Sitter spaces, have also attracted some interest
recently, due to the phenomenon of black hole anti-evaporation \cite{bousso}.
Besides, these black holes could be relevant to the proposed duality
between the large $N$ limit of Euclidean four-dimensional $U(N)$
super-Yang-Mills theory and the so-called type IIB$^*$ string theory in
de~Sitter space \cite{hull}.

In five dimensions, extremal solutions of Einstein-Maxwell gravity were
discussed in \cite{gkltt}\footnote{Cf.~also \cite{gauntlett,mirjam} for
further discussions of five-dimensional
rotating black holes.}. These solution are supersymmetric in the sense that
they preserve half of the supersymmetry when viewed as bosonic solutions of
the pure $N=2$ ungauged supergravity theory. The metric for these black
holes is of the Tangherlini form \cite{frt}. Later, extreme black holes for
$N=2$ ungauged supergravity coupled to abelian vector multiplets were
considered in \cite{sabra1,sabra2,sabra3}. An important feature of these
black holes is that, for those with non-singular horizons, the entropy can
be expressed in terms of the extremum of the central charge. Black hole
solutions of gauged supergravity theory have also been the subject of many
recent studies. It is known from the results of \cite{bcs1} that
supersymmetric solutions of this theory (with negative cosmological
constant) have naked singularities, and therefore one need to study
non-extremal or possibly supersymmetric rotating solutions\footnote{%
In four dimensions, it has indeed been shown that rotating BPS solutions can
have an event horizon \cite{caldo}. For further studies of supersymmetric
$AdS_4$ black holes cf.~\cite{romans}.}. The non-extremal generalizations
of the solutions considered in \cite{bcs1} were studied in \cite{bcs2}.
General rotating charged solutions in five-dimensional (anti)de~Sitter
spaces are not known yet. It is our purpose in this work to study these
solutions, and as a first step in this direction, we will be mainly
concerned with the solutions of five-dimensional Einstein-Maxwell gravity
with Chern-Simons term and with a positive or negative cosmological constant.

Anti-de~Sitter rotating solutions without charge in five dimensions have
been recently found by Hawking et al.~\cite{hawking1}. These solutions break
all supersymmetries when viewed as solutions of $N=2$ gauged supergravity,
unless they are massless. In order to have non-trivial rotating solutions
preserving some supersymmetry, it is thus necessary to include gauge fields.

This work is organized as follows. In the next section, we present our
rotating solution to Einstein-Maxwell gravity with cosmological constant. In
section \ref{physprop}, we study some physical properties of the charged
rotating de~Sitter black holes, like horizons, Hawking temperature and
entropy. Furthermore, we find a Smarr formula, from which a first law of
black hole mechanics can be derived. In section \ref{multicent} we express
our solution in so-called cosmological coordinates, which will allow us to
find also multi-centered rotating charged de~Sitter black holes in five
dimensions. We conclude with some final remarks.

\section{Rotating charged black hole solutions in (anti)de~Sitter space}

In this section we will present new rotating charged solutions in five
dimensional (anti)-de~Sitter spaces. Consider the Einstein-Maxwell theory in
five dimensions with cosmological constant. This theory contains the
graviton and one abelian gauge field. The action is given by \footnote{%
Our action is related to that of \cite{london} by the rescaling of the gauge
fields of \cite{london} by $\frac{1}{2\sqrt{3}}$ and multiplying the action
by an overall factor of 2. We use the metric $\eta _{ab}=(-,+,+,+,+)$.} 
\begin{equation}
S=\frac{1}{4\pi G_{5}}\int d^{5}x\,e\left( \frac{1}{2}R-\Lambda -\frac{1}{24}%
F_{\mu \nu }F^{\mu \nu }+\frac{e^{-1}}{216}\epsilon ^{\mu \nu \rho \sigma
\lambda }F_{\mu \nu }F_{\rho \sigma }A_{\lambda }\right) ,  \label{action}
\end{equation}
where $\mu ,\nu $ are spacetime indices, $R$ is the scalar curvature,
$F_{\mu \nu }=\partial _{\mu }A_{\nu }-\partial _{\nu }A_{\mu }$ denotes the
abelian field-strength tensor, and $e=\sqrt{-g}$ is the determinant of the
f\"{u}nfbein $e_{\mu }^{a}$. $G_{5}$ denotes the five-dimensional Newton
constant, and $\Lambda =-6g^{2}$ is the cosmological constant. In the
context of supergravity, the gauging of the theory enforces the addition of
a cosmological constant with $g$ being the coupling constant of the
gravitino to the gauge field. This means that for negative $\Lambda ,$ the
action (\ref{action}) can be regarded as the bosonic part of the gauged
supersymmetric five-dimensional $N=2$ supergravity theory without vector
multiplets.

The Einstein and gauge field equations of motion derived from (\ref{action})
read

\begin{eqnarray}
R_{\mu\nu} &=& F_{\mu\nu}^2-\frac16 g_{\mu\nu}F^2+4g^2 g_{\mu\nu},
\label{einstein} \\
\partial_{\nu}(eg^{\mu\rho}g^{\nu\sigma}F_{\rho\sigma}) &=& \frac{1}{12}
\epsilon^{\mu\nu\rho\sigma\kappa}F_{\nu\rho}F_{\sigma\kappa},  \label{gauge}
\end{eqnarray}

where $F^2=\frac16 F_{\mu\nu}F^{\mu\nu}$ and $F_{\mu\nu}^2=\frac16
g^{\rho\lambda}F_{\mu\rho}F_{\nu\lambda}$.

The above equations admit the (anti) de~Sitter vacuum as a solution which
(for negative $\Lambda$) is maximally supersymmetric. This solution has zero
gauge fields, and the metric can be represented by\footnote{%
For $g$ real (negative cosmological constant) we have the anti-de~Sitter
solution and for imaginary $g$ (positive cosmological constant) we get
de~Sitter space.}

\begin{equation}
ds^2=-(1+g^2r^2)dt^2+\frac{1}{1+g^2r^2}dr^2+r^2d\Omega^2,  \label{ds1}
\end{equation}
where $d\Omega^2$ denotes the standard metric on the unit three-sphere.

From the results of \cite{london, bcs1} we know that the equations (\ref
{einstein}), (\ref{gauge}) admit BPS black hole solutions (with naked
singularities for negative cosmological constant) asymptotic to (\ref{ds1}),
and breaking half of the supersymmetry. These solutions are given by

\begin{equation}
ds^{2}=-H^{-2}(1+g^{2}r^{2}H^{3})dt^{2}+\frac{H}{1+g^{2}r^{2}H^{3}}%
dr^{2}+r^{2}Hd\Omega ^{2},  \label{non}
\end{equation}
\begin{equation}
A_{t}=3H^{-1},
\end{equation}

where $H=(1+\frac{q}{r^{2}})$, and $q$ is essentially the electric charge.

For $g=0$, the metric (\ref{non}) reduces
to the extreme Reissner-Nordstr\"{o}m black hole of five-dimensional
Einstein-Maxwell gravity without cosmological constant, given by

\begin{equation}
ds^2 = -H^{-2}dt^2+Hd\vec{x}^2,  \label{RN}
\end{equation}
where $d\vec{x}^2$ is the flat line element in four dimensions. (\ref{RN})
is of course the simplest example of the generalizations of
Majumdar-Papapetrou solutions \cite{mp} to five dimensions, given by

\begin{eqnarray*}
ds^2 &=& -\Omega^{-2}dt^2+\Omega d\vec{x}^2, \\
A_t &=& 3\Omega^{-1}, \quad \Omega = 1+\sum_{j=1}^N\frac{q_j}{|\vec{x}-\vec{x%
}_j|^2}.
\end{eqnarray*}

By introducing the coordinate $R^2=r^2+q$, the solution (\ref{non}) can be
written in the form

\begin{equation}
ds^2 = -g^2R^2dt^2-\frac{1}{R^4}(R^2-q)^2dt^2+\frac{dR^2}{U(R)}+R^2d\Omega^2,
\label{nonrot}
\end{equation}
where 
\begin{equation}
U(R)=\left(1-\frac{q}{R^2}\right)^2+g^2R^2.
\end{equation}

We now present our solution which is the rotating generalization of (\ref
{nonrot}). The equations of motion (\ref{einstein}) and (\ref{gauge}) are
satisfied for the metric

\begin{equation}
ds^2 = -g^2R^2dt^2 - \frac{1}{R^4}((R^2-q)dt - \alpha\sin^2\theta d\phi +
\alpha\cos^2\theta d\psi)^2 + \frac{dR^2}{V(R)} + R^2d\Omega^2,
\label{metrdS}
\end{equation}
where 
\begin{equation}
V(R)=\left(1-\frac{q}{R^2}\right)^2+g^2R^2-\frac{g^2\alpha^2}{R^4},
\label{V}
\end{equation}

and the gauge fields

\begin{equation}
A_{\phi} = -3\frac{\alpha\sin^2\theta}{R^2},\quad A_{\psi} =3\frac{\alpha
\cos^2\theta}{R^2},\quad A_t = 3\left(1-\frac{q}{R^2}\right).
\end{equation}

Notice in the above solution that $g$ can be real and pure imaginary,
corresponding to anti-de~Sitter or de~Sitter solutions.

\section{Physical properties}

\label{physprop}

In the following we will concentrate on de~Sitter charged rotating black
holes and their physical properties, and leave the study of anti-de~Sitter
solutions for a future publication.

\subsection{Horizons}

Horizons occur whenever $V(R)=0$. This implies that\footnote{%
As we are interested in the de~Sitter case, we substitute $g\to ig$ in (\ref
{V}).}

\begin{equation}
\left(1-\frac{q}{R^2}\right)^2-g^2R^2+\frac{g^2\alpha^2}{R^4} = 0.
\end{equation}
This equation admits a solution representing a cosmological horizon for all
parameter values. To study the issue of the existence of black hole event
horizons, we first define the dimensionless parameters

\begin{equation}
a \equiv \alpha g^3, \quad \varrho \equiv q g^2.
\end{equation}

One finds then that for $0 < 6\varrho < 1$ and

\begin{equation}
a_-^2(\varrho) < a^2 < a_+^2(\varrho),
\end{equation}

where

\begin{equation}
a_{\pm}^2(\varrho) = -(\varrho^2 - \frac 23 \varrho + \frac{2}{27}) \pm 
\frac{2}{27}(1 - 6\varrho)^{3/2},
\end{equation}

one has a black hole with (inner) Cauchy horizon $R_-$ and (outer) event
horizon $R_+$. Clearly, there is also a cosmological horizon at $R=R_c > R_+$%
. For $a^2 = a_-^2(\varrho)$, the event and cosmological horizons coalesce,
the resulting metric is then similar to the Nariai solution \cite{nariai}.
For $a^2 = a_+^2(\varrho)$, the Cauchy and event horizons coalesce, so that
one has an extremal black hole in de~Sitter space. Another interesting case
appears for $6\varrho = 6\sqrt 3 a = 1$. The function $V(R)$ then reads

\begin{equation}
V(R) = g^2R^2\left(\frac{1}{3g^2R^2} - 1\right)^3,
\end{equation}

which implies that we have in this case an "ultracold" cosmological horizon.

\subsection{Temperature and entropy}

We now wish to compute the Hawking and cosmological temperatures. To this
end, we write the metric (\ref{metrdS}) in the canonical (ADM) form

\begin{equation}
ds^2 = -N^2dt^2 + \sigma_{mn}(dx^m + N^mdt)(dx^n + N^ndt),
\end{equation}

with the lapse function

\begin{equation}
N^2 = \frac{V(R)}{1-\frac{\alpha^2}{R^6}},
\end{equation}

and the shift vector

\begin{equation}
N^{\phi} = -N^{\psi} = \frac{\alpha(R^2-q)}{R^6-\alpha^2}.
\end{equation}

Consider now the analytical continuation $t \to -i\tau$, which yields the
"quasi-Euclidean" section

\begin{equation}
ds^2 = N^2d\tau^2 + \sigma_{mn}(dx^m - iN^md\tau)(dx^n - iN^nd\tau).
\label{quasieucl}
\end{equation}

We assume to have either an event or a cosmological horizon at $R=R_H$, and
use the expansion

\begin{equation}
V(R) = V^{\prime}(R_H)(R-R_H)
\end{equation}

near $R=R_H$. Defining the new coordinate

\begin{equation}
\tilde{R} = 2\sqrt{\frac{R-R_H}{V^{\prime}(R_H)}},
\end{equation}

(\ref{quasieucl}) can be written near $R=R_{H}$ as

\begin{equation}
ds^2 = d\tilde{R}^2 + \left(\frac{V^{\prime}(R_H)}{2\sqrt{1-\frac{\alpha^2} {%
R_H^6}}} \right)^2\tilde{R}^2d\tau^2 + R_H^2d\theta^2 + \sigma_{ij} (dx^i -
iN^id\tau)(dx^j - iN^jd\tau),  \label{quasieuclnear}
\end{equation}

where $i,j = \phi,\psi$. From (\ref{quasieuclnear}), we see that the period
of $\tau$ must be

\begin{equation}
\beta = \frac 1T = \frac{4\pi}{|V^{\prime}(R_H)|}\sqrt{1-\frac{\alpha^2}{%
R_H^6}}
\end{equation}

in order to avoid conical singularities. The angular velocities of the
horizon read

\begin{equation}
\Omega^{\phi}_H = -\Omega^{\psi}_H = -N^{\phi}_H.
\end{equation}

Note that in the case of zero cosmological constant, the angular velocities
of the event horizon vanish \cite{gauntlett}, whereas for $g \neq 0$ the
horizon rotates.

The Bekenstein-Hawking entropy of the de~Sitter black holes under
consideration is given by

\begin{equation}
S_{BH} = \frac{A_H}{4G_5} = \frac{\pi^2}{2G_5}(R_+^6 - \alpha^2)^{1/2},
\end{equation}

which reduces in the case $g=0$ to the result found in \cite{sabra2}.

\subsection{Smarr formula}

To obtain a Smarr-type formula (from which a first law of black hole
mechanics can be deduced), we proceed along the lines of \cite{gibbhawk},
where an analogous calculation for the four-dimensional
Kerr-Newman-de~Sitter black hole was performed. We start from the Killing
identity

\begin{equation}
\nabla_{\mu}\nabla_{\nu}K^{\mu} = R_{\nu\rho}K^{\rho} = (F_{\nu\rho}^2
-\frac16 g_{\nu\rho}F^2)K^{\rho} + 4g^2K_{\nu},  \label{killingid}
\end{equation}

where $K^{\mu }$ is a Killing vector, $\nabla _{\left( \nu \right.
}K_{\left. \mu \right) }=0$. Note that in the second step we used the
Einstein equation of motion (\ref{einstein}). Now integrate (\ref{killingid}%
) on a spacelike hypersurface $\Sigma _{t}$ from the black hole horizon $%
R_{+}$ to the cosmological horizon $R_{c}$. On using Gauss' law this gives

\begin{equation}
\int_{\partial\Sigma_t}\nabla_{\nu}K_{\mu}d\Sigma^{\mu\nu} = 4g^2
\int_{\Sigma_t}K_{\nu}d\Sigma^{\nu} + \int_{\Sigma_t}(F_{\nu\rho}^2 -
\frac16g_{\nu\rho}F^2)K^{\rho}d\Sigma^{\nu},  \label{B1}
\end{equation}

where the boundary $\partial\Sigma_t$ consists of the intersection of $%
\Sigma_t$ with the black-hole and the cosmological horizon,

\begin{equation}
\partial\Sigma_t = S^3(R_+) \cup S^3(R_c).
\end{equation}

In a first step, we apply (\ref{B1}) to the Killing vectors $\partial_{\phi}$
and $\partial_{\psi}$, which we denote by $\tilde{K}^i$, $i = \phi,\psi$.
Using $\tilde{K}^i_{\nu}d\Sigma^{\nu} = 0$, we get

\begin{equation}
\frac{1}{8\pi}\int_{S^3(R_+)}\nabla_{\nu}\tilde{K}^i_{\mu}d\Sigma^{\mu\nu} + 
\frac{1}{8\pi}\int_{S^3(R_c)}\nabla_{\nu}\tilde{K}^i_{\mu}d\Sigma^{\mu\nu} =
\int_{\Sigma_t}T_{\nu\rho}\tilde{K}^{i\rho}d\Sigma^{\nu},  \label{C}
\end{equation}

where

\begin{equation}
T_{\nu\rho} = \frac{1}{8\pi}\left(F_{\nu\rho}^2-\frac14g_{\nu\rho}F^2\right)
\end{equation}

is the stress-energy tensor of the vector field. One can interpret the
right-hand side of (\ref{C}) as the angular momentum of the matter between
the two horizons. Thus one can regard the second term on the left-hand side
of (\ref{C}) as being the total angular momentum, $J_c^i$, contained in the
cosmological horizon, and the first term on the left-hand side as the
negative of the angular momentum $J_{BH}^i$ of the black hole (cf.~also
discussion in \cite{gibbhawk}).\newline
In a second step, we apply (\ref{B1}) to the Killing vector $K=\partial _{t}$%
. This yields

\begin{eqnarray}
\lefteqn{\frac{1}{4\pi}\int_{S^3(R_+)}\nabla_{\nu}K_{\mu}d\Sigma^{\mu\nu} +%
\frac{1}{4\pi}\int_{S^3(R_c)}\nabla_{\nu}K_{\mu}d\Sigma^{\mu\nu}=}  \notag \\
&&\frac{g^2}{\pi}\int_{\Sigma_t}K_{\nu}d\Sigma^{\nu}+2\int_{\Sigma_t}
\left(T_{\nu\rho}-\frac13Tg_{\nu\rho}\right)K^{\rho}d\Sigma^{\nu}.  \label{D}
\end{eqnarray}

One can regard the terms on the right-hand side of Eq.~(\ref{D}) as
representing respectively the contribution of the cosmological constant and
the contribution of the matter kinetic energy to the mass within the
cosmological horizon. We therefore identify the second term on the left-hand
side as the total mass $M_c$ within the cosmological horizon, and the first
term as the negative of the black hole mass $M_{BH}$. As in Ref.~\cite{4laws}%
, the latter can be rewritten by expressing $K=\partial_t$ in terms of the
null generator

\begin{equation}
l=\partial _{t}+\Omega _{H}^{\phi }\partial _{\phi }+\Omega _{H}^{\psi
}\partial _{\psi }=\partial _{t}+\Omega _{H}^{i}\tilde{K}^{i}
\end{equation}

of the black hole event horizon. Using the definition of the surface gravity 
$\kappa$, which, by the zeroth law, is constant on the horizon, one obtains

\begin{equation}
M_{BH} = \frac{\kappa A_H}{4\pi} + 2\Omega^i_HJ^i_{BH}.
\end{equation}

One therefore gets the Smarr-type formula

\begin{eqnarray}
M_c &=& \frac{\kappa A_H}{4\pi}+2\Omega_H^{\phi}J_{BH}^{\phi}+2\Omega_H^{%
\psi} J_{BH}^{\psi}+\frac{g^2}{\pi}\int_{\Sigma_t}K_{\nu}d\Sigma^{\nu} 
\notag \\
&& + 2\int_{\Sigma_t}\left(T_{\nu\rho}-\frac13 Tg_{\nu\rho}\right)K^{\rho}
d\Sigma^{\nu}.
\end{eqnarray}

\section{Cosmological multi-centered solutions}

\label{multicent}

Multi-centered solutions of five-dimensional gravity with a cosmological
constant have been discussed in \cite{london} and more recently in \cite{jw1}
for the cases with vector multiplets. Multi-centered charged solutions in
four dimensional de Sitter space were considered in \cite{kt}.

In this section we generalize these solutions and find multi-centered
rotating charged five-dimensional de~Sitter black holes. In order to find
these solutions, we recast the 5d de~Sitter metric in the so-called
cosmological coordinates, in which the metric appears similar to Minkowski
space but with the Euclidean part multiplied by an exponential depending on
time and the cosmological constant. This metric is given by 
\begin{equation}
ds^2=-dt^2+e^{2gt}(dr^2+r^2d\Omega^2).  \label{dsn}
\end{equation}

The metric of $dS_5$ given in (\ref{ds1}) can be obtained from (\ref{dsn})
by performing the change of variables

\begin{equation}
r^{\prime}=re^{gt},\quad dt=dt^{\prime}-\frac{gr^{\prime}}{1-g^2 {r^{\prime}}%
^2}dr^{\prime}.  \notag
\end{equation}

For our rotating charged black hole solution (\ref{metrdS}) we find that it
can be written (after substituting $g\to ig$) in the form

\begin{eqnarray}
ds^2 &=& -H^{-2}(dt-e^{-2gt}\frac{\alpha}{r^2}\sin^2\theta d\phi + e^{-2gt} 
\frac{\alpha}{r^2}\cos^2\theta d\psi)^2 + He^{2gt}(dr^2+r^2d\Omega^2),
\label{rotmetrcosm} \\
A_{\phi} &=& -\frac{3\alpha\sin^2\theta}{r^2H}e^{-2gt}, \quad
A_{\psi} = \frac{3\alpha\cos^2\theta}{r^2H}e^{-2gt}, \quad
A_t = 3H^{-1}, \notag
\end{eqnarray}

where we defined 
\begin{equation*}
H \equiv 1+\frac{q}{r^2}e^{-2gt}.
\end{equation*}
(\ref{rotmetrcosm}) is related to the metric of (\ref{metrdS}) by a change
of variables according to

\begin{eqnarray*}
r^{\prime } &=&re^{gt},\quad dt=dt^{\prime }+f(r^{\prime })dr^{\prime }, \\
d\phi &=&d\phi ^{\prime }+h(r^{\prime })dr^{\prime },\quad d\psi =d\psi
^{\prime }-h(r^{\prime })dr^{\prime },
\end{eqnarray*}

where

\begin{eqnarray*}
f(r^{\prime }) &=&\frac{-gr^{\prime }\left( (1+\frac{q}{{r^{\prime }}^{2}}%
)^{3}-\frac{\alpha ^{2}}{{r^{\prime }}^{6}}\right) }{1-g^{2}{r^{\prime }}%
^{2}(1+\frac{q}{{r^{\prime }}^{2}})^{3}+\frac{g^{2}\alpha ^{2}}{{r^{\prime }}%
^{4}}}, \\
h(r^{\prime }) &=&\frac{\alpha g}{{r^{\prime }}^{3}\left( 1-g^{2}{r^{\prime }%
}^{2}(1+\frac{q}{{r^{\prime }}^{2}})^{3}+\frac{g^{2}\alpha ^{2}}{{r^{\prime }%
}^{4}}\right) },
\end{eqnarray*}

and subsequently setting $R^2={r^{\prime}}^2+q$ and dropping the primes.
As for the unrotating
solutions \cite{london,jw1}, the harmonic function can be chosen to be 
\begin{equation}
H=1+\sum_{j=1}^N\frac{q_j}{|\vec{x}-\vec{x}_j|^2}e^{-2gt}.  \label{harm}
\end{equation}

In this case one obtains a multi-centered solution representing an arbitrary
number of charged rotating black holes in a space-time with positive
cosmological constant.

\section{Final remarks}

In this work we have obtained new solutions of Einstein-Maxwell gravity with
Chern-Simons term and cosmological constant. Some physical properties of the
rotating charged de~Sitter solutions have also been studied. We also found
the coordinate transformation which recast our de Sitter solution in the
so-called cosmological coordinates. In the cosmological coordinates one
finds the generalization of our solution to the multi-centered case.

\bigskip 

As in the unrotating case, we note here that our rotating anti-de Sitter
solution is BPS and breaks half of the supersymmetry of the $N=2$, $d=5$
gauged supergravity theory. This means that our solution solves Killing
spinor equation,

\begin{equation}
\left( \mathcal{D}_{\mu }+\frac{i}{24}(\Gamma _{\mu }{}^{\nu \rho }-4\delta
_{\mu }{}^{\nu }\Gamma ^{\rho })F_{\nu \rho }+\frac{1}{2}g\Gamma _{\mu }-%
\frac{i}{2}gA_{\mu }\right) \epsilon =0. \label{gst}
\end{equation}

Here $\mathcal{D}_{\mu }$ is the gauge and gravitationally covariant
derivative. The Killing spinor is the same as in the unrotated
case \cite{london,jw1},

\begin{equation}
\epsilon =H^{-\frac12}e^{igt}\epsilon_0,
\end{equation}

where $H$ is given by (\ref{harm}), and $\epsilon _{0}$ is a constant spinor
satisfying the constraint $\Gamma_0\epsilon_0=i\epsilon_0$.
Clearly, the de Sitter solution admits a Killing spinor which can be
obtained by replacing $g$ with $ig$ in the above expressions.

We should also note here that the rotating solution in anti-de~Sitter space
has a horizon unlike the static solutions which have a naked singularity.
The supersymmetry properties of the charged rotating AdS black holes as well
as their relevance to the AdS/CFT correspondence are currently under study.

\section*{Acknowledgements}
The part of this work due to D.~K.~was partially
supported by the European Commission RTN program
HPRN-CT-2000-00131, in
which D.~K.~is associated to the University of Torino.

\end{document}